\documentclass{aastex}
\usepackage[twocolumn]{emulateapj5}

\renewenvironment{figure}{\vspace{2mm}\noindent%
\refstepcounter{figure}%
\renewcommand{\caption}[1]{\par\footnotesize\textsc{Fig.~\thefigure.---}\
  ##1\par}}%

\newcommand{\fov}{field of view}
\newcommand{\V}{{\cal V}}
\newcommand{\dx}{\vec{\Delta x}}
\newcommand{\thetaE}{\theta_{\rm E}}
\newcommand{\tE}{t_{\rm E}}
\newcommand{\umin}{u_{\rm min}}

\begin{document}
\title{Astrometric imaging of crowded stellar fields with only two 
{\sl SIM} pointings}

\author{Neal Dalal \& Kim Griest}
\affil{Physics Dept., University of California, San Diego, CA 92093}
\email{endall@astrophys.ucsd.edu, griest@astrophys.ucsd.edu}

\begin{abstract}
The Space Interferometry Mission (SIM) will observe sources in crowded
fields.  Recent work has shown that source crowding can induce
significant positional errors in SIM's astrometric measurements, even
for targets many magnitudes brighter than all other crowding sources.
Here we investigate whether the spectral decomposition of the fringe
pattern may be used to disentangle the overlapping fringes from
multiple blended sources, effectively by performing synthesis imaging
with two baselines.  We find that spectrally dispersed fringes enable
SIM to identify and localize a limited number of field sources quite
robustly, thereby removing their effect from SIM astrometry and
reducing astrometry errors to near photon noise levels.  We simulate
SIM measurements of the LMC, and show that (a) SIM astrometry will not
be corrupted by blending and (b) extremely precise imaging of mildly
crowded fields may be performed using only two orthogonal baseline
orientations, allowing microarcsecond positional measurements.  We
lastly illustrate the method's potential with the example of
astrometric microlensing, showing that SIM's mass and distance
measurements of lenses will be untainted by crowding.
\end{abstract}

\section{Introduction}
The Space Interferometry Mission (SIM) will perform astrometry with
unprecedented precision, measuring positions at the microarcsecond
level.
This will allow numerous astrophysical experiments to be performed
that previously were impossible.  Even though its 10 meter baseline
allows resolution only of $\sim 10$ mas, SIM achieves microarcsecond 
astrometric precision by centroiding fringes to better than 1 part in
a thousand.  If SIM can meet its technical specifications, the primary
source of error for single star astrometry would be photon noise,
which can be beat down as $1/\sqrt{N}$ until the systematic error
floor is reached.  For multiple stars in the field of view, however,
other errors may be incurred.  For example, the extra flux of photons
from field sources gives additional photon noise, which increases the
amount of on-source integration time to achieve a given astrometric
precision.  Another source of error, which cannot be reduced by
further integration time, is the distortion in the measured fringe
pattern caused by the superposition of fringes from multiple
incoherent sources.  Since SIM achieves its revolutionary precision by
fitting the observed fringe pattern in exquisite detail, even small
scale deviations from the single star fringe pattern can lead to large
astrometric errors.  Several possible SIM experiments, such as
microlensing, globular cluster dynamics, or stellar motions in
external galaxies like M31, require SIM astrometry in crowded
fields, and so it is necessary to determine whether or not crowding
will corrupt SIM measurements irretrievably.

One might imagine that SIM astrometry of targets many magnitudes
brighter than other sources in the \fov\ would be unaffected by 
the presence of those sources.
However, \cite{jayadev} have recently shown that source confusion
in crowded fields can lead to significant
errors ($\sim$ few $\mu$as) in individual position measurements, even
when the confusing field sources are just a few percent of the
target's brightness.  Such
errors are comparable to those expected from photon noise and
systematic error, and could pose a serious threat to astrometric
measurements in crowded fields if uncorrected.  
Brighter sources, such as binary companions or highly luminous
field stars $\gtrsim 5\%$ in brightness, could completely 
destroy SIM astrometry.  To correct these
measurements, one would like to know the positions and brightnesses of
all the important sources in the \fov\ -- that is, one would like an image.
A possibility is to obtain, prior to SIM observations,
either space telescope or ground based adaptive optics (AO)
snapshots of the field.
Unfortunately, as we show below, the rough positions obtainable ($\sim
40$ mas) from such images are of insufficient precision to rectify
SIM astrometry.  Another alternative is to perform synthesis imaging
with SIM.  In analogy to radio interferometry, one might orient SIM
along numerous baselines, using multiple siderostat pairs to fill in
the $(u,v)$ plane \citep{allen}.  Note, however, that a possible SIM
design currently under consideration involves just one 10 meter baseline.
A third alternative, which we discuss here, is to use SIM in its 
astrometry mode.  If the sources all lie several fringes away from each 
other then the $(u,v)$ coverage obtained from just
two baselines, using multiple frequency channels, is sufficient to
disentangle the sources from each other and measure all of their
positions to few $\mu$as precision.  This is a marked increase in
astrometric precision relative to that possible from SIM imaging, $\sim
250\,\mu$as \citep{allen}, and thus allows significantly
improved proper motion measurements.
We show that, for the currently expected
characteristics of the SIM satellite, and for typical levels of crowding
in the LMC, Bulge, and the outer regions of 
globular clusters, these two orthogonal
baselines are adequate to achieve excellent astrometric or proper
motion precision.  Note that similar ideas have been used previously
in simulations of closure phase measurements \citep{schloerb} and in
visibility measurements of the binary star Capella \citep{koechlin}.

\begin{figure}
\noindent\centerline{
\includegraphics[width=0.43\textwidth]{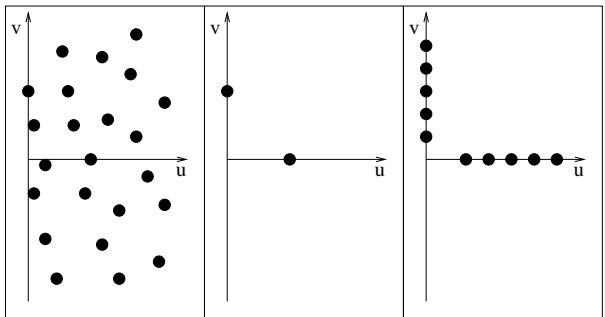}}
\caption{Schematic depictions of $(u,v)$ coverage for (a) synthesis
imaging with many baseline orientations, 
(b) astrometry with two baseline orientations, and (c) astrometry with
two baselines and channeled spectrum.
Only half of the $(u,v)$ plane is illustrated since
the brightness distribution is real; redundant points are not
plotted. 
\label{uv}}
\end{figure}

To see why just two baselines are sufficient, consider a synthesis
imaging measurement of several point sources.  This involves measuring
the complex visibility $\V$ at numerous points in the $(u,v)$ plane,
where $u=B_x/\lambda$, $v=B_y/\lambda$ are spatial frequencies, and
$B_x$ and $B_y$ are the projections of the baseline separation along
the $x,y$ axes respectively.  The visibility $\V(u,v)$ and brightness
distribution $F(x,y)$ are related by Fourier transformation,
e.g. $\V(u,v)=\int dxdy F(x,y) \exp(-2\pi i[ux+vy])$, via the
van Cittert-Zernike theorem \citep[see for example][Chapter
2]{lawson}.  For an arbitrary brightness distribution $F(x,y)$
sufficiently many $u,v$ points must be sampled to permit accurate
Fourier inversion of the visibility.  A typical astrometry measurement
relies upon just two baseline orientations, which would appear
insufficient to localize more than one source position ($x$ and $y$
coordinates).  Note, however, that each frequency channel at a given
baseline orientation constitutes an independent $u,v$ measurement,
since $u=kB_x/2\pi$, $v=kB_y/2\pi$ \citep{schloerb}.  
The multifrequency $u,v$ coverage
from two orthogonal SIM baselines is limited to points on the $u,v$
axes (see Figure \ref{uv}), and is clearly insufficient to reconstruct
an arbitrary brightness distribution.  For {\it point sources}, on the
other hand, this coverage suffices.  Consider $F(x,y)=\sum_i
F_i(k)\delta(x-x_i)\delta(y-y_i)$ corresponding to a sum of point
sources, each with spectrum $F_i(k)$.  Taking for now each $F_i$
independent of $k$ gives the measured 
visibility as $\V(u,v)=\sum_iF_i\exp(-2\pi i[x_iu+y_iv])$.  Now,
instead of the usual 2D Fourier transform to recover the brightness
distribution, consider a single 1D Fourier transform, say along the
$u$ direction.  This gives $\sum_if_i(v)\delta(x-x_i)$, and similarly
for the 1D transform along the $v$ direction.  Clearly, the sources'
$x$ coordinates are measurable just by sampling the visibility at
multiple $u$ along a line of constant $v$, and similarly for the $y$
positions.  By placing each source on a color-magnitude diagram (CMD)
using the $F_i(k)$,
the $x$ sources can be matched up with the $y$ sources, giving the
unique 2D coordinates of each bright source in the field \citep{dyck}.  
If the
pairing is not obvious from the CMD, then a third baseline orientation
is required to match up the $x$ and $y$ sources, but note that this
need be done only once, at any time.  In theory, therefore, the two
orthogonal SIM baselines required to perform single source astrometry
are all that are necessary to perform imaging of point sources.  In
practice, finite bandwidth, Nyquist sampling and photon noise will
limit point source resolution.  We explore these effects in
the next section.

\vspace{-5mm}\section{Fitting dispersed fringes}
The above procedure of Fourier transforming along the $u$ or $v$
direction is equivalent to taking the spectrally dispersed 1D fringe
pattern for each orthogonal baseline orientation, and Fourier
transforming with respect to frequency $k$.  This is the same 
procedure that is often used for single source phase and group delay 
estimation \citep[Chapter 8]{lawson}.  Let $F(k)$ equal the
product of source spectrum, throughput, and detector response.  
Note that since CCD's count numbers of photons, not power, $F$
has units of numbers.  The
interferometer response at frequency $k$ to a source offset from the
phase center by an angle $\theta$ along the baseline $B$ is
$\case{F(k)}{2}(1+\cos kd)=\case{F(k)}{2}(1+\cos kB\theta)$ where $d$
is the physical delay.  Fourier transforming with respect to $k$ gives
$[2\tilde F(l)+\tilde F(l-d) + \tilde F(l+d)]/4$, with $l$ the Fourier
conjugate to $k$ and $\tilde F(l)$ the Fourier transform of $F(k)$.  
In most cases, $F(k)$ is slowly varying, and so its Fourier transform
has power mainly at low frequencies.  For example, for $F(k)$=const,
$\tilde F(l)=\delta(l)$ and the Fourier transform of the fringe
pattern becomes $[2\delta(l)+\delta(l-d)+\delta(l+d)]/4$.  
Likewise, for a rectangular bandpass, we would have a sinc function.
For simplicity, we will focus on rectangular $F(k)$, although the
generalization is clear.

\begin{figure}
\begin{centering}\plotone{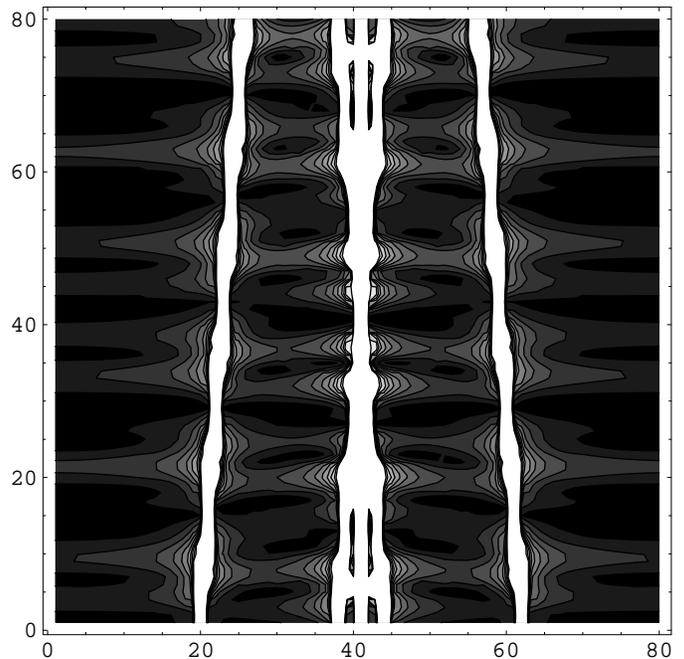}\end{centering}
\caption{Fourier transform of the channeled spectrum with two 
equally bright sources, one at the phase center and the other at 250 mas.
The abscissa is the Fourier conjugate to wavenumber, the ordinate
is delay, and the numbers along the axes label pixels.  
Note that sources appear as lines in the Fourier transform,
which cross at the position (delay) of the respective source.
Additional sources should appear as additional, distinct lines.  This
figure neglects photon noise, which can hide the signal from faint sources.
\label{crosses}}
\end{figure}

As expected, each source in the \fov\ appears as a narrow peak
in the Fourier transform space (actually a pair since the brightness
is a real quantity).  Also note that half of the power lies
at zero frequency, due to the 1/2 which offsets the average fringe
intensity from zero.  Figure \ref{crosses} illustrates the pattern for
two point sources.  

\vspace{2mm}
\vspace{2mm}

Now, assume that SIM uses $n_c$ frequency channels.  For simplicity,
we take each channel to have the same frequency width.  
This assumption allows the use of the
FFT, otherwise a more general discrete Fourier transform 
or Lomb-Scargle periodogram would be
used.  With this assumption, the Nyquist theorem tells us that
each bin in Fourier space has a size equaling the white-light 
coherence envelope size 
$\lambda_{\rm max}\lambda_{\rm min}/[B(\lambda_{\rm max}-\lambda_{\rm min})]$
which for a rectangular $0.4-1.0\,\mu$m bandpass and 10 meter baseline is
roughly 13.75 mas.  This makes sense, since SIM should be able to
resolve sources separated by more than one coherence envelope size.  
Clearly, decreasing the effective bandwidth should diminish SIM's
ability to resolve point sources.  Varying the number of channels
should affect only aliasing.  With fewer, broader channels, a source
near the edge of the \fov\ has reduced fringe contrast and would be
mistaken for the combination of a fainter source near the middle of
the FOV and uniform surface brightness.  This could corrupt
astrometric measurements of sources towards the edge of the \fov.

We have simulated SIM astrometric measurements of crowded fields with
the above considerations.  We assume that SIM scans repeatedly over a
small range in delay, corresponding to the central fringe size at the
mean frequency, which for a 10 m baseline and $0.4-1.0\,\mu$m bandpass
is 11.8 mas.  We divide this range into $n_d$ bins 1 mas large, and
integrate the fringe pattern $f(k,d)=F_i(1+\cos kd)/2$ over 
$n_d\times n_c$ total bins, where $n_c=64$ is the number of wavelength
channels spanning the bandpass.  We sum the contributions for all
sources in the field of view, and approximate photon noise by adding
to each bin a random number of photons drawn from a Gaussian with
$\sigma=\sqrt{N}$.  We neglect all instrumental noise, such as read
noise, dark current, thermal drift, etc., basically assuming that SIM
is a perfect astrometer.  These additional errors may become important
for the faintest sources.  To simulate SIM astrometry and photometry,
we fit the resulting fringe pattern to the form 
$$
N_{\rm model}=N_0 + \sum_i N_i \frac{1+\cos kB(\theta-\theta_i)}{2}.
$$
The number of sources, their positions and brightnesses are first
estimated using the Fourier transform method described above.
Starting with
these initial values we fit by minimizing the error function 
$\chi^2=\sum_{i,j}(N_{ij,{\rm data}}-N_{ij,{\rm model}})^2/N_{ij,{\rm data}}$.
The fitted parameters are each source's brightness $N_i$ and position
$\theta_i$, as well as uniform surface brightness $N_0$.  This assumes
that the product of the source spectrum, throughput, and detector
response is constant with respect to wavelength, but it is not
difficult to generalize to include simple spectral dependence.  

An estimate of the centroiding precision possible by such fringe
fitting may be computed using the minimum variance bound,
hereafter MVB \citep{mvb}.
When fitting data to a function $f(x)$ with normalization and one
additional parameter $x_0$, the minimum variance expected in the
fitted $x_0$ is 
$$
{\rm var}(x_0)\equiv\sigma_{x_0}^2=\frac{1}{N}\left[\left\langle\left(
\frac{d\ln f}{dx_0}\right)^2\right\rangle-\left\langle\frac{d\ln f}
{dx_0}\right\rangle^2\right]^{-1}
$$
where $\langle g\rangle\equiv(\int fg\,dx)/(\int f\,dx)$.  Using the
MVB formula, and taking $N$ total photons from a single point source,
we expect centroiding errors $\sigma_\theta=1.87 N^{-1/2}$ mas for
scans covering 11 mas and the wavelength range $0.4-1\,\mu$m.  Using the
white-light fringe pattern ($n_c$=1) we expect
$\sigma_\theta=2.91N^{-1/2}$ mas.  So for $N=340,000$ photons we
expect 3.2 $\mu$as and 5 $\mu$as errors for the spectrally dispersed and
white-light fringe pattern, respectively.  This analytic estimate
compares well with a Monte Carlo calculation of the expected
astrometric errors using the above fringe fitting procedure.  Using
bins of size 1 mas ($n_d\approx 11$) gives errors of 3.8 $\mu$as and
5 $\mu$as, in agreement with the MVB estimate for white-light fringes
and somewhat larger than the MVB estimate for the spectrally dispersed
fringe pattern.  Increasing the number of bins to $\sim 100$ gives
Monte Carlo estimates in agreement with the MVB estimate.  One might
conclude from this that it is always better to centroid using
spectrally dispersed fringes rather than broadband, white-light
fringes since the errors are smaller for a given number of photons.
However, keep in mind that we have neglected additional errors
such as read noise, and there may be a reduction in throughput when
the fringes are spectrally dispersed.

\section{Fitting multiple sources}
Having confirmed that our Monte Carlo errors agree with analytic
estimates for single sources, we now explore errors for multiple sources.
As mentioned earlier, additional sources in the \fov\ can increase
astrometry errors by contributing extra photon noise, and by
distorting the fringe shape.  Fitting for multiple sources cannot
remove photon noise, but it can mitigate the distortion arising from
superimposed fringes.  To ease the comparison with unblended
centroiding as described in the previous section, we set the
on-source integration time to equal the time necessary to accumulate
340,000 photons from the science target.  In terms of a 
signal-to-noise ratio 
(S/N), this holds fixed the signal while letting the noise
vary.  When gauging how well fitting can remove the effects of
blending, we should compare to a case with the same ratio of signal
(photons from science target) to noise (square root of total number of
photons).  For example, consider centroiding a single star in the
presence of uniform surface brightness that contributes an equal
number of photons as the star.  Since the S/N is smaller by a
factor of $\sqrt{2}$ relative to the isolated star case, the
centroiding errors will be greater by a factor of $\sqrt{2}$.

As our first example, we consider measurements of a bright
microlensing event in the LMC.  
We place a $V=19$ target at the phase
center, and randomly place field stars in the 800 mas FOV.
The number of field stars, and their luminosity function (LF), are
chosen to match the crowding and LF of the LMC derived by the MACHO
project \citep{thor}.  On average, MACHO finds 10-16 field stars
within 1-2'' of each observed star.  To be conservative, we have used
16 field stars in all of our simulations.
The average brightness
of a star drawn from this LF is approximately 1.4\% the brightness of
a $V=19$ star, so by adding 16 of them, we add approximately 22\% more
photons.  We thus expect a roughly 10\% increase in astrometry errors from
photon noise alone.  Since we normalize the on-source observation time
to give us 3.8 $\mu$as photon noise errors from the $V=19$ source
alone, we expect 4.2 $\mu$as errors from photon noise.  Any additional
errors will be due to source confusion \citep{jayadev}.

\begin{figure*}
\centerline{
\includegraphics[width=0.47\textwidth]{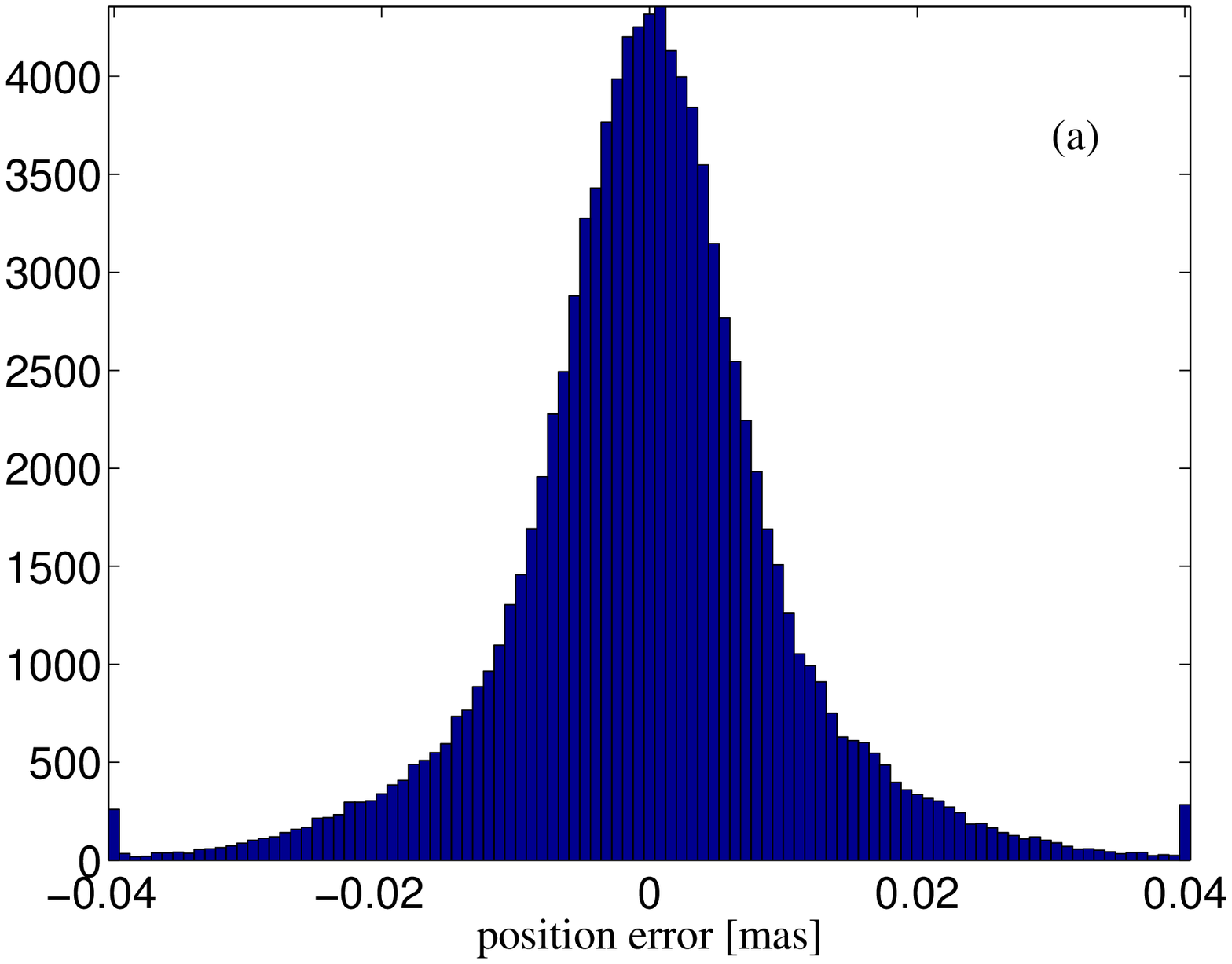}
\includegraphics[width=0.47\textwidth]{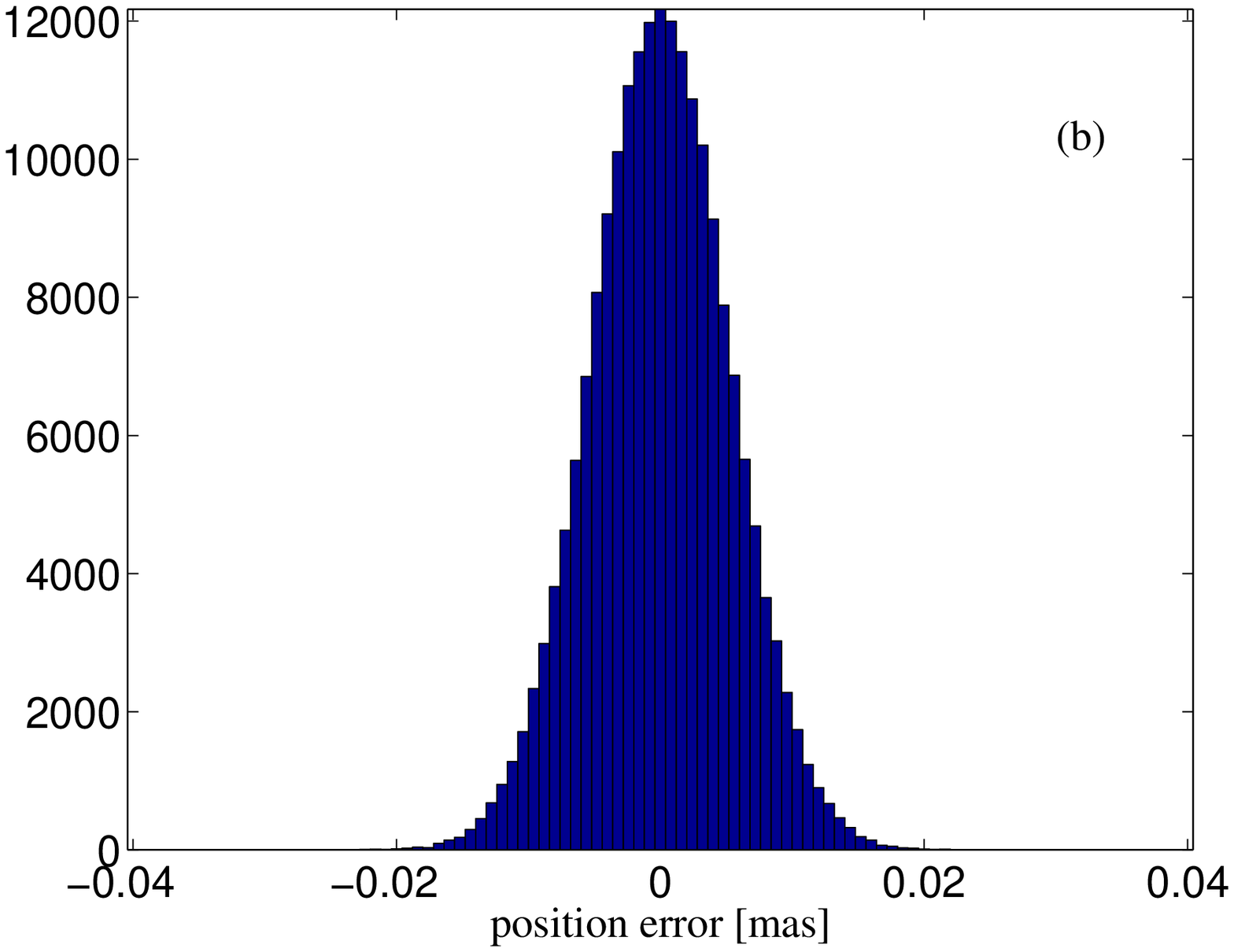}}
%\begin{centering}
%\plottwo{1star+fuzzhistogram.eps}{restrictedhistogram.eps}
%\end{centering}
% the data for panel a is in results/1star+fuzzhistogram.dat
% the data for panel b is in results/restrictedhistogram.dat
\caption{Histogram showing position residuals for multiple source
fitting to $V=19$ source and 16 stars drawn from LMC luminosity
function.  The first panel shows the histogram for random placement of
stars.  The peaks at the edge of the plot are
due to events falling outside the histogram range.
The second panel shows the histogram for the same calculation,
restricting sources to lie more than 25 mas from the target at the
phase center.  See the text for more details.
\label{lmc_1}}
\end{figure*}

We follow the same
fitting procedure described above, allowing for multiple sources at
different positions and with different brightnesses.  Note, however,
that faint sources may not be detectable above photon noise, so we
will only be able to remove the brightest sources.  Since in this
example we are only interested in the measurement errors for our
single target, we only report measurement errors for the $V=19$
target.  Panel (a) of Figure \ref{lmc_1} shows the histogram of
position residuals for this simulation.  Overall, the standard
deviation of this error distribution is $\sigma_\theta=10.5\,\mu$as,
much larger than the 4.2 $\mu$as
errors expected from photon noise.  Note, however, that
the histogram may not be well described by a single ``average'' error
value.  We find that the error distribution
is characterized by a narrow core superimposed on a slowly decaying 
tail of outliers.  This histogram is quite similar to
that found by \citet{jayadev}, although for different reasons.
In their simulations of white-light astrometry in crowded fields in
M31, the LMC and Galactic Bulge,
the long tail was due to the occasional bright field star.
Here, in contrast, bright field stars are easily identified and
removed using dispersed fringes.  The long
tail here is due to the occasional star landing nearby our target and
becoming unresolved from it.  To demonstrate this point, 
panel (b) of Figure \ref{lmc_1} shows the histogram for the same
simulation, but here we restrict field stars to lie more than 25 mas
from our target.  Note that the long tails have vanished, and that the
distribution's standard deviation has significantly diminished, 
to 5.3 $\mu$as, near to the value expected from photon noise.
Clearly, almost all of the 10.5 $\mu$as errors in this case arise from
unresolved or marginally unresolved stars falling close to the target.
\cite{jayadev} have shown that these quite significant errors largely
cancel for proper motion measurements, and so for many SIM
applications they are not a worry.  
For microlensing, however, confusion errors will be a problem in
crowded fields.  This is because microlensed stars 
%are not exceptionally bright, they 
undergo different proper motions than other
stars in the field (even binary companions), and their brightnesses
vary during the course of the lensing event.  Therefore, the
cancellation of errors that occurs for typical proper motion
measurements will not occur here, and we expect proper motion errors
of the same magnitude as the large position errors found above.  
Fortunately, as we show below, it
is still possible to disentangle the blend effect using the complete
set of position measurements.  Thus, even in this most badly affected
case, it is still possible to recover accurate
masses and distances from astrometric microlensing events.

\begin{figure}
\begin{centering}\plotone{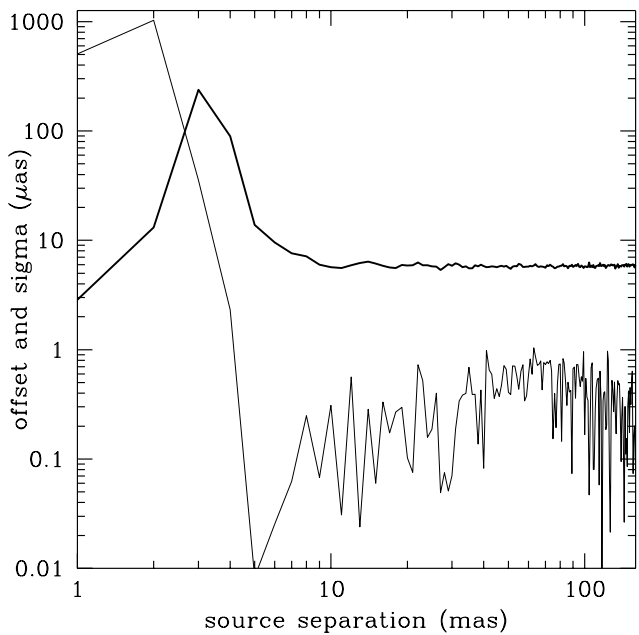}\end{centering}
\caption{Fit residuals for astrometric measurements of a single star,
with an equally bright blend in the field of view, as a function of
distance to the blend.  The thin curve is the average offset of the
fit source position from the input position and the thick curve is the
standard deviation of those offsets.
\label{2starposition}}
\end{figure}

Generally, blend stars in SIM observations will be far fainter than
the intended science targets, which are chosen for their brightness.
However, it is still worthwhile to consider what effects comparably
bright sources would have.  To quantify this, we simulate SIM
measurements of two equally bright stars in the field of view.  
Again we normalize the on-source integration time to give 340,000
photons from the science target.  Photon noise alone should thus give
us errors of $\sqrt{2}\cdot 3.8\,\mu{\rm as}=5.4\,\mu$as.
We plot the results of our simulations in Figure \ref{2starposition}.
We can distinguish three kinds of blends from this figure.  First,
note that bright blends falling outside the central fringe are
resolved by SIM and can be removed.  The offsets due to such sources
are consistent with zero, and the dispersions ($\approx 6\,\mu$as) are
not much larger than what is expected from photon noise alone.
Clearly, such blends pose no threat to SIM astrometry using
spectrally dispersed fringes.  The average
offset is not zero, however, for sources falling within the central
fringe.  Sources at distances $\leq$ 2 mas are completely unresolved
by SIM.  As expected, then, the measured centroid falls at the center
of light, which is 0.5 mas for an equally bright blend at 1 mas, and 1
mas for an equally bright blend at 2 mas.  Note that although the
offset is large, the standard deviation is small.  This is just what
we expect; two equally bright sources at 0 and 1 mas should appear to
SIM to be a source twice as bright located at 0.5 mas, and so the
errors should be smaller by a factor of $2^{-1/2}$ compared to single
source astrometry.  The only blend positions giving significant errors
are those that are marginally unresolved, i.e. falling barely inside
the central fringe, at separations of 3-4 mas.  There would seem to be
no way to handle such blends with a single baseline orientation,
however an additional pointing along a different baseline orientation
could resolve the two sources.

\begin{figure*}
\begin{centering}
%\plottwo{10sources_hist1.eps}{10sources_errs.eps}
%\plotone{imagingerrors.eps}
\includegraphics[width=0.51\textwidth]{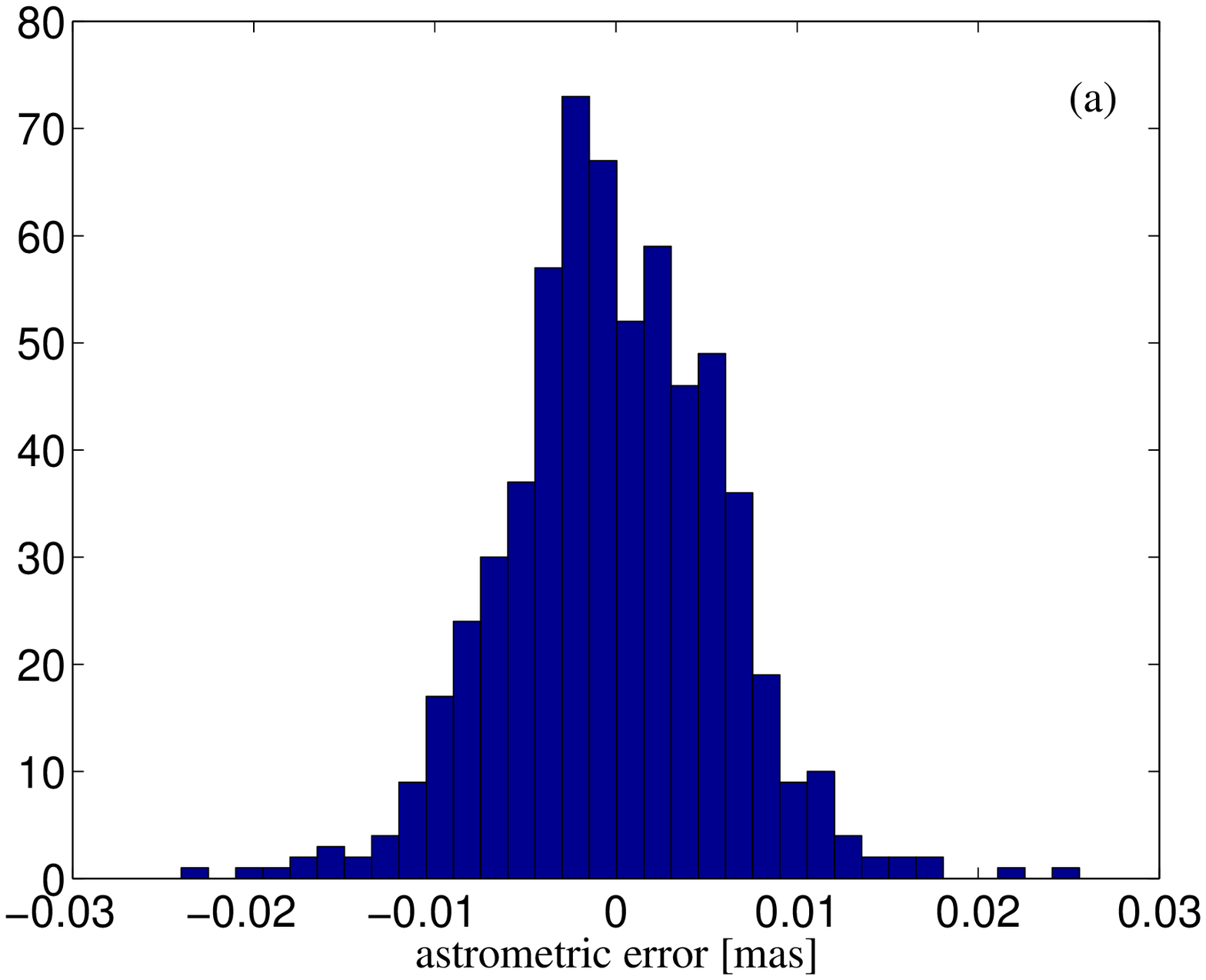}
\includegraphics[width=0.43\textwidth]{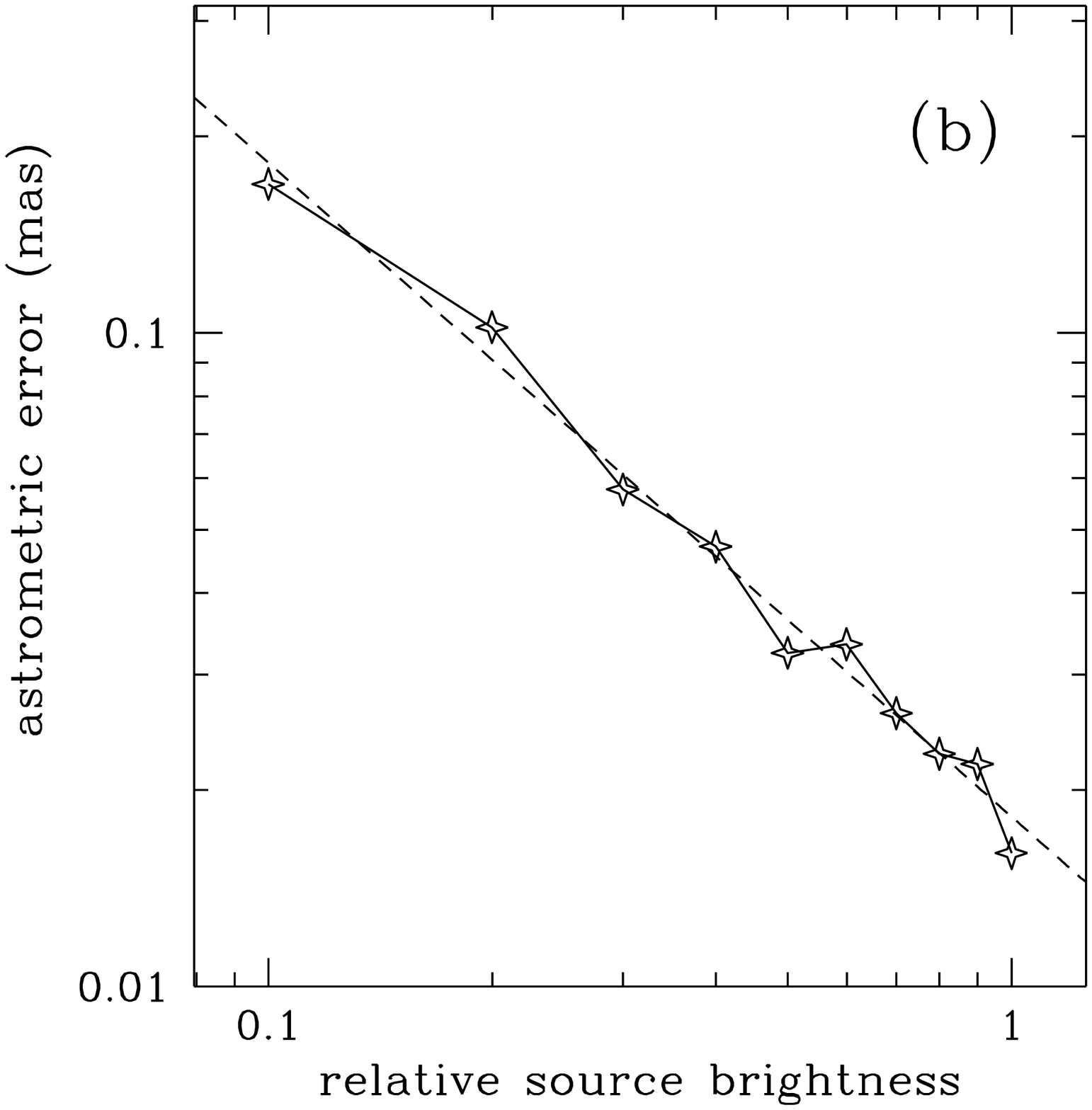}
\end{centering}
\caption{Simulated astrometric imaging errors for 10 sources.
Panel (a) show the error histogram from a simulation
of imaging of 10 equally bright sources in the 800 mas \fov.  
The standard deviation of
this histogram is $\sigma=6\,\mu$as.  Panel (b) shows the scaling of
error with brightness for a simulation of 10 sources with varying
brightnesses.  Plotted points are errors as a function of brightness
relative to the brightest star in the field.  The dashed line is not a
fit, but shows the expected scaling: error inversely proportional to
brightness.
\label{10stars}}
\end{figure*}

It is apparent from Figure \ref{2starposition} that the vast majority
of bright field stars will pose no threat to SIM astrometry.
The fact that the fit residuals are so small, comparable to photon
noise, means that fitting can reliably remove the fringes from
blending sources in the \fov.  Since we are able to recover $\mu$as
astrometry even in the presence of blends, this must mean that we are
able to fit the blend positions to microarcseconds themselves.  Since
we are doing this for all important blends in the \fov, we are
basically imaging the \fov.  It is therefore interesting to explore
the imaging capabilities of SIM using just the spectrally dispersed
fringes measured along two orthogonal baselines.  From the discussion
above, we expect that we can get reliable positions as long as the
sources are separated by more than one white-light coherence envelope.
In terms of the Fourier transform space, we expect to measure reliable
positions for sources separated by more than one pixel.  With $n_c$
wavelength channels, we have $n_c/2$ distinct pixels in the Fourier
conjugate variable since the brightness is real.  We have assumed 64
channels, so we have 32 pixels.  Assuming that sources should be
separated by 3 pixels for reliable positions means that we expect that
for up to 10 sources in the FOV, we can measure positions reliably.
To test this, we simulated SIM astrometric measurements of 10 stars 
in the field of view.  We first simulated 10 stars of equal
brightness; the results are shown in panel (a) of Figure \ref{10stars}.
We find that the astrometric errors scale like $t^{-1/2}$ where $t$ is
the on-source integration time, so they are mainly due to
photon noise.

\noindent 
For the simulation plotted in panel (a), the
astrometric errors were $\sigma=6\ \mu$as.  In panel (b), we plot the
scaling of errors with brightness.  Using the
above arguments on S/N, we expect the errors to scale inversely with
each star's brightness, and this is verified in Figure \ref{10stars}(b).
Clearly, far smaller errors are attainable using dispersed fringes
than the $\sim 250\,\mu$as errors expected from synthesis 
imaging of many
baseline orientations (although keep in mind that we have neglected
errors such as thermal drift).  Thus, it is possible that highly
precise astrometric imaging may be achieved using only two SIM pointings.

\section{Application: Astrometric microlensing}
One possible application of SIM astrometry in crowded fields is
microlensing.  There are numerous astrophysical applications of SIM
observations of microlensing events 
\citep{pac,boden98,planet,simphoto,nearby}, 
most of which entail crowded fields.  For
ordinary proper motion, \citet{jayadev} have shown that astrometric
errors arising from source confusion largely cancel.  This is because
any offset caused by additional sources remains basically constant as
the target executes its proper motion, and cancels when the relative
motion of the target is measured.  However, this is not true for
microlensing.  The reason for this is that microlensing changes the
target's apparent brightness while also causing the apparent proper
motion.  The astrometric offset caused by blending, therefore, does
not remain constant and will not cancel.  Thus, we might expect that
blending could seriously damage microlensing measurements, rendering
SIM observations useless in crowded fields.  Fortunately, the results
discussed in previous sections indicate that blending by bright stars
should not corrupt SIM astrometry, except when those sources fall
within the white-light coherence envelope of the target star.  However,
binary companions to lensed stars could easily fall inside the central
fringe.  Since fringe fitting could never detect such stars, 
just as fringe fitting cannot resolve the two microlensed images, our
concerns about crowding could still hold.  Fortunately, as we describe
below, even sources falling within the central fringe may be detected
and removed, not from individual SIM measurements, but by using the
entire set of observations of the microlensing event.

As discussed in the previous section, field sources may be detected
and essentially removed as long as they lie outside the central fringe
of the target.  In fact, even marginally resolved sources falling
barely within the central fringe may be detected and removed if bright
enough, by the distortion they cause in the fringe shape.  The only
sources undetectable are those which do not distort the fringe shape,
but merely shift its centroid.  These sources, lying well inside the
central fringe, shift the fringe centroid to the center of light of
the target and blend, and add all of their light to the target.  What
SIM would detect would be a star of brightness equaling the sum of
the target and blend's brightness, located at their center of light.
For this reason, multiple blends falling inside the central fringe are
equivalent to a single blend, and so we consider only the single blend
case.  Such blends affect the microlensing lightcurve and astrometric
motion in a simple way \citep{han}, and we show that SIM observations may
be corrected without degeneracy to extract the desired measurables.

It is first necessary to review astrometric microlensing; see
\citet{boden98} for more details.  Microlensing of a single source
by a single lens gives
two images of the source.  If the source is located at position
${\vec x_s}=\thetaE{\vec u}$ relative to the lens, then the two
images appear at positions with magnifications
$${\vec x_{1,2}}=\thetaE(u\pm\sqrt{u^2+4})/2$$
$$A_{1,2}=\frac{u^2+2}{2u\sqrt{u^2+4}}\pm\frac{1}{2}.$$
Here, $\thetaE$ is the angular Einstein radius, $u=|{\vec u}|$ is the
lens-source separation in units of $\thetaE$ (not the $u$ from the
$u,v$ plane), and positions are relative to the lens position.
SIM cannot resolve these images, so in the absence of blending it
detects a source with position ${\vec x}={\vec x_s} + \dx$ 
and magnification $A$ given by
$$\dx=\thetaE\frac{\vec u}{u^2+2}, \qquad
A=\frac{u^2+2}{u\sqrt{u^2+4}}.$$
Note that $\dx$ traces out an ellipse.

\begin{figure}
\centerline{\includegraphics[width=0.45\textwidth]{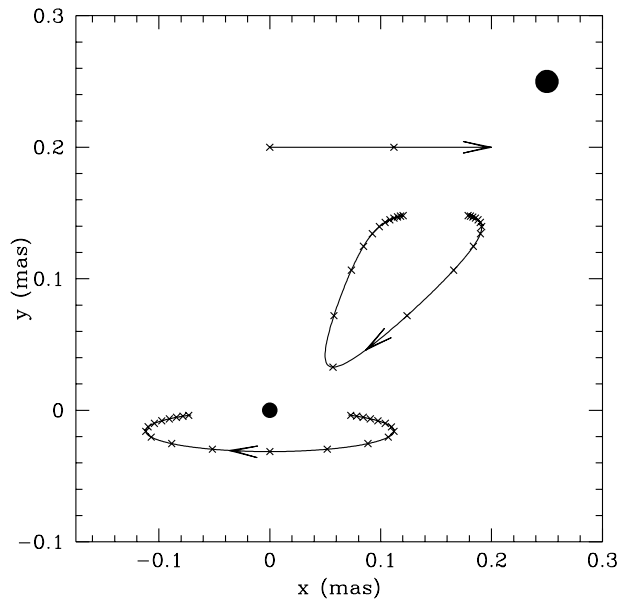}}
\caption{
Microlensing source and motion of lens and image center-of-light.
The large dot shows the location of the blend star, the small dot the
location of the target star.  The straight arrow shows the 
location and relative motion of the lens near the peak of the event, 
with x's placed every week.  $\tE=20$ days and $\thetaE=0.32$ mas.  
The elliptical curve asymptoting to the target source shows
the unblended astrometric motion of the 
image center of light, while the
irregular curve is the observed motion of the center of light including
the blend star light.  Arrows show the direction of motion.
\label{blendexample}}
\end{figure}

It is now easy to see how unresolved blends affect this.  If the
lensed source has unlensed brightness $F_s$, and the blend has brightness
$F_b$ and position ${\vec x_b}$, then SIM detects a source at position
${\vec x}_{\rm obs}$ with brightness $F_{\rm tot}$ given by
$$
F_{\rm tot}=AF_s+F_b, \quad {\vec x}_{\rm obs}=
\frac{AF_s({\vec x_s} + \dx) +F_b{\vec x_b}}{F_{\rm tot}}
$$
An example of this is given in Figure \ref{blendexample}.
From the illustration, it may appear that blending would ruin SIM
astrometry, since the observed motion bears little resemblance to the
unblended motion.  However, our
simulations indicate that it will be possible to recover the unblended
motion.  
This is largely because SIM's exquisite photometry
(recall $N\gtrsim 340,000$) permits the nondegenerate decomposition of the
photometric lightcurve into lensed and unlensed (blend) light.  That
is, $F_s$, $F_b$, and $A$ may be measured from the photometry alone.
Given the blend fraction, we now show that
it is easy to fit the astrometric data to recover the microlensing
parameters $\thetaE$, $\tE\equiv|\dot{\vec u}|^{-1}$, etc., which are
then used to determine the lens mass and distance.

\begin{figure}
\begin{centering}
\includegraphics[width=0.45\textwidth]{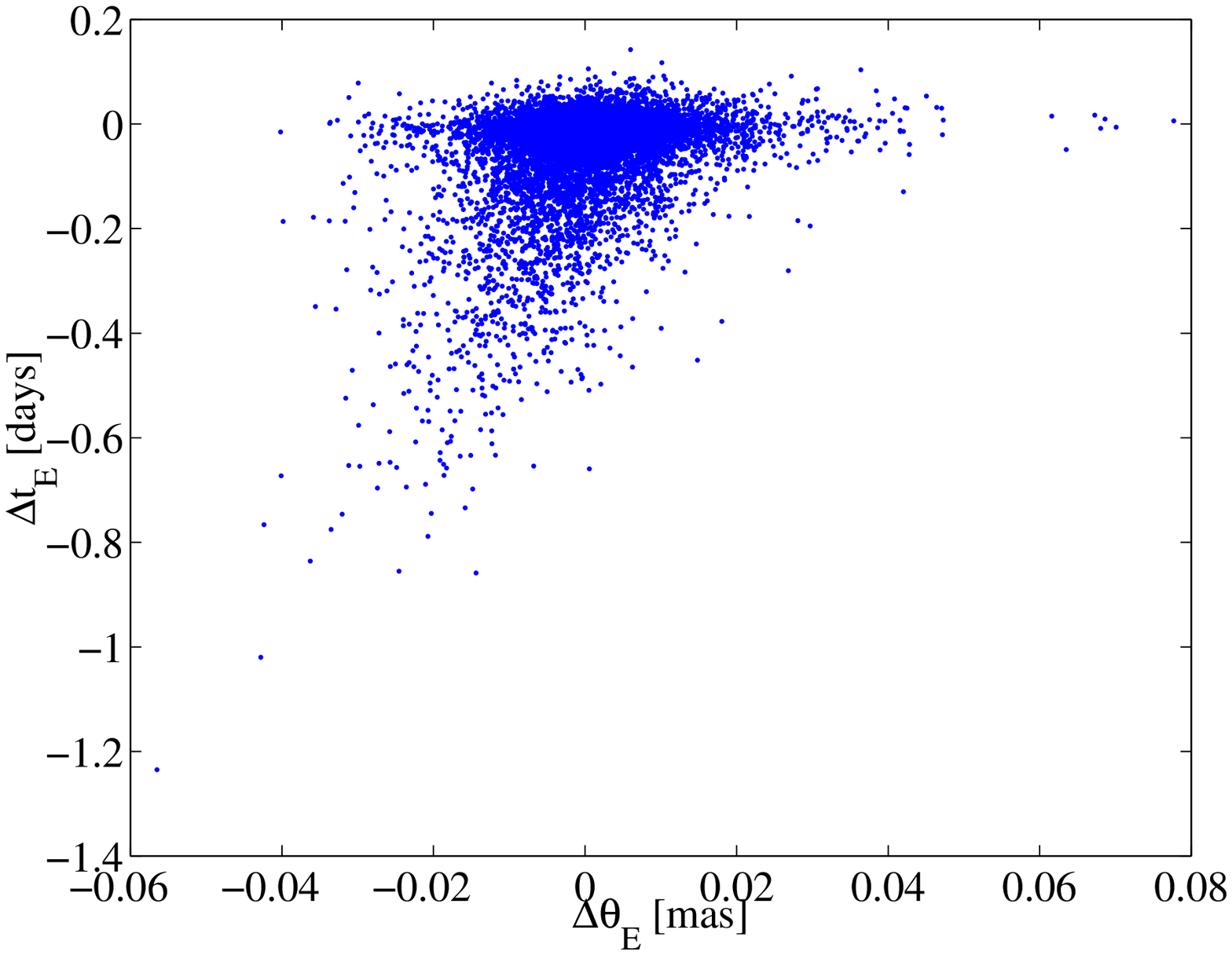}\\
\includegraphics[width=0.45\textwidth]{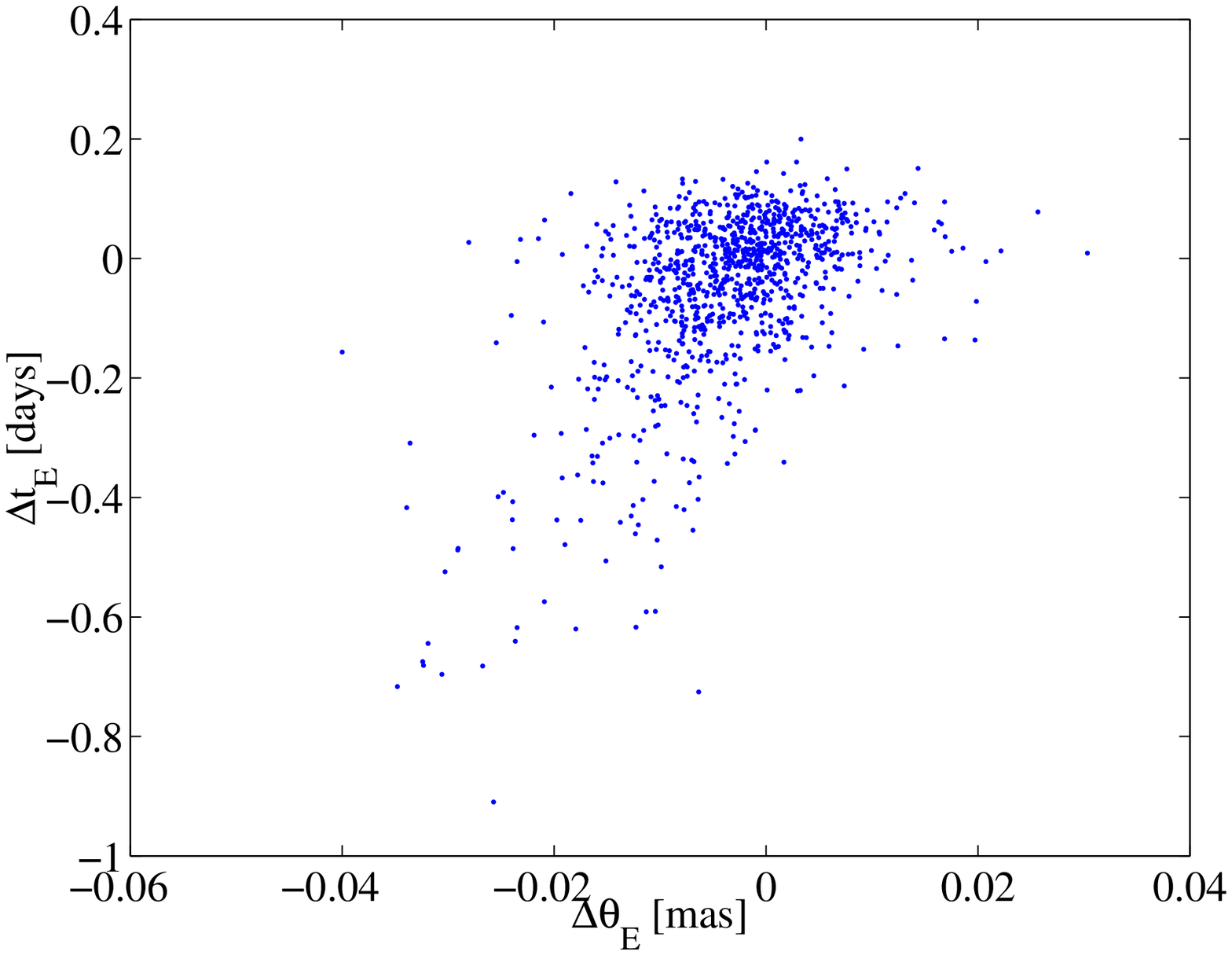}\\
\includegraphics[width=0.45\textwidth]{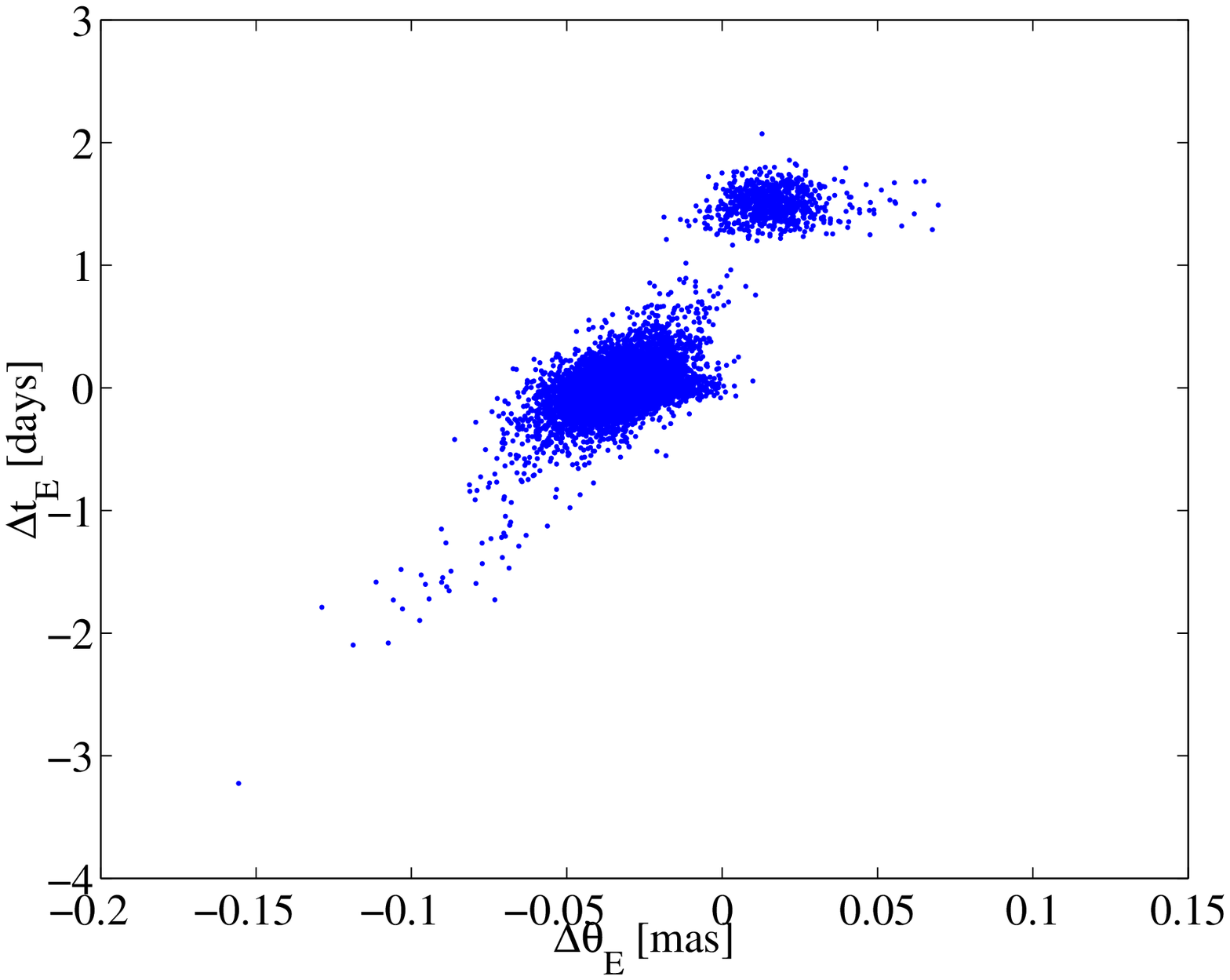}\\
\end{centering}
\caption{Fit residuals for microlensing event.  Each point in the
scatter plot represents a different simulated event; plotted are the
residuals in fitted $\thetaE$ and $\tE$.  10,000 simulations of
unblended microlensing events are plotted in the first panel; the
standard deviations are $\sigma_\theta=0.009$ mas and $\sigma_t=0.1$
days, for input values $\thetaE=0.32$ mas and $\tE=20$ days.
In the second panel, we plot 1000 simulations of microlensing events
with a $V=19$ target near the phase center and 16 additional stars
drawn from LMC luminosity function placed randomly in FOV.  Here, the
standard deviations are $\sigma_\theta=0.008$ mas and $\sigma_t=0.14$
days for the same input parameters as the unblended case.
The last panel plots errors for 10,000 simulations of microlensing
events with 16 field stars, but with white-light fringe fitting.
Here, $\thetaE$ is systematically underestimated by 0.03 mas (roughly
10\%), with a dispersion of 0.017 mas, while $\tE$ has a dispersion of
0.44 days.
\label{microres}}
\end{figure}

The blended microlensing event is parametrized by 11 quantities: the
above mentioned $\thetaE$, $\tE$, $\vec x_s$, $\vec x_b$, $F_s$,
and $F_b$, as well as the normalized impact parameter $\umin$,
time of closest approach $t_0$, and relative proper motion direction
$\phi$.  For simplicity we assume that the same blend brightness $F_b$
applies for both the $x$ and $y$ astrometric measurements.  This will
not be true in general (a field star whose projected position relative
to the target happens to fall inside the central fringe under one
orientation need not fall inside the central fringe along the
orthogonal orientation).  However, it is trivial to generalize our
treatment to include two different blends for the $x$ and $y$
measurements.  We have also assumed that the blend's proper motion
is negligible for simplicity, but again it should not be difficult to
generalize to moving blends.
Now, the photometry alone gives us $t_0, \tE,
\umin, F_s,$ and $F_b$.  The remaining 6 parameters
are to be derived from
the astrometry.  First note that the unblended astrometric motion
along the relative proper motion vector is antisymmetric in time with
respect to $t_0$, while the unblended motion perpendicular to it is
symmetric (see Figure \ref{blendexample}).  
The blend position and brightness are constant, and
therefore symmetric.  So if we multiply $\vec x_{\rm obs}$ by $F_{\rm
tot}$, and antisymmetrize in time with respect to $t_0$, then the
blend parameters drop out, and the resulting motion is along the
unblended proper motion vector.  So we immediately get $\phi$.  We can
estimate $\thetaE$ from $\tE$, $\umin$ and the velocity
$\dot{\vec x}$ at peak.  Again, note that the blend parameters make a
constant contribution to the product $F_{\rm tot}\vec x_{\rm obs}$, so
by taking the difference of this quantity evaluated at two different
times, the blend parameters drop out.  We can thus estimate the
unlensed source position $\vec x_s$ by using the estimates for the
other parameters, and the change in $F_{\rm tot}\vec x_{\rm obs}$ from
peak to the end of the event.  Now we have estimates for all
parameters except the blend position $\vec x_b$, which we can guess by
comparing the measured $F_{\rm tot}\vec x_{\rm obs}$ with the
contributions from the other terms.  Thus, we can simply and
accurately guess values for all of the microlensing parameters, and
use these as initial estimates for fitting routines.  This is
also true if there are different blend fractions for the two baseline
orientations.

To quantify the above statements, we have simulated SIM observations
of microlensing events.  Each of our simulated events has
$\thetaE=0.32$ mas and $\tE=20$ days.  The other parameters, such as
$\umin$, $\phi$ and $\vec x_s$ were picked randomly.  We measure 25
points during the event, evenly spaced in time, starting when the
apparent magnification exceeds a threshold $A>A_{\rm th}=1.5$, and
lasting $5\tE$ from the trigger.  This is probably not an optimal
observing strategy, but it is not our intention here to optimize the
strategy.  Each astrometric measurement is made following the
procedure described in previous sections -- the fringe
pattern is simulated and fitted to recover the source position and
brightness.  We assume that the $x$ and $y$ measurements occur at the
same time, although this assumption has no effect.  If SIM measures
the orthogonal baselines at different times, say to minimize its
overall slewing, no information should be lost.  Even if the baselines
are not perfectly orthogonal, measurements of the
microlensing parameters should not be seriously degraded.

To measure how badly crowding affects microlensing measurements, we
first see how well the parameters may be recovered for a single,
unblended source.  To save on computation time, we used white-light
fringe measurements, which as noted above should not be much worse
than spectrally dispersed measurements for unblended sources.  Now, to
measure the lens mass and distance, the parameters of interest are
$\thetaE$, $\tE$, and $\phi$.  In the first panel of Figure
\ref{microres} we plot fit residuals for the above observing strategy.
Note that these fits utilize both SIM astrometric and photometric
measurements of the microlensing events.  We find that $\thetaE$ can
be measured to 2.84\%, and $\tE$ to 0.5\%, for input values of
$\thetaE=0.32$ mas and $\tE=20$ days.

Now, we add blend stars to see how badly they affect microlensing mass
and distance measurements.  Here, we clearly need to use the full
multifrequency fringe fit to resolve bright blend stars.  From the
individual fringe fits, which gave errors $\sim 10\,\mu$as (twice as
large as the unblended errors) we might think that the errors here
would be twice as large.  We plot the fit residuals in the second panel
of Figure \ref{microres}.  We obtain errors $\sigma_\theta=0.008$ mas,
$\sigma_\phi=0.0235$, and $\sigma_t=0.142$ days.  Normalized to
$\thetaE=0.32$ mas and $\tE=20$ days, these are fractional errors of
2.5\%, 2.4\%, and 0.71\% respectively, as good as or better than the
unblended white light errors!  Again, the reason that these errors are
not twice as big as the unblended errors is that the relatively large
single measurement residuals mainly arose from unresolved sources
falling inside the target's central fringe, which as we discussed may
easily be removed.

Another important question to ask is how badly we would do without
the spectral decomposition of the fringe pattern.  To address this
question, we repeated the above simulations, but this time used
only white-light fringes.  The resulting errors are plotted in Figure
\ref{microres}.  We find that the error distribution is
bimodal, and neither blob is centered on the correct answer!  The bulk
of the fits underestimate $\thetaE$ by roughly 10\%, with a scatter of
$\sim\pm 5\%$, although a subset of events give somewhat overestimated
$\thetaE$, about $5\%$ too large, with a scatter of $\sim\pm 3\%$.  
This ``more correct'' subset, however, overestimates $\tE$ by roughly
$8\%$.  So if we were to fit a white light measurement of a
microlensing event in the LMC, we could be reasonably certain that the
derived masses and distances would be incorrect by about 20\%.
While $20\%$ distances would certainly be good enough to resolve the 
question of the location of the LMC lenses, it is clearly preferable
to utilize the channeled spectrum when possible.

\begin{figure}
\centerline{\includegraphics[width=0.46\textwidth]{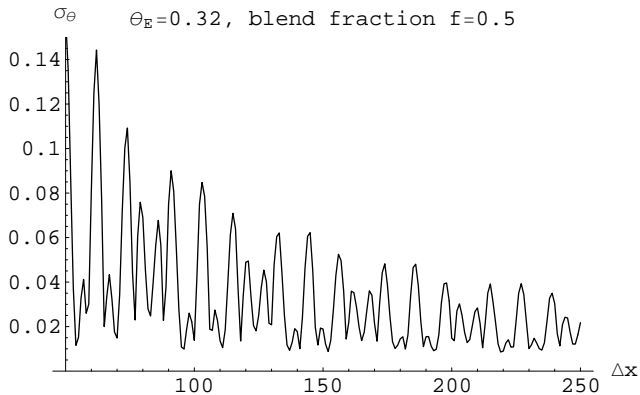}}
\caption{Fit residuals for microlensing event with an equally bright
blend in FOV, fitting to the white-light fringe pattern.  We
only plot for distances large compared to the fringe size; 
closer separations give much larger errors.  $\sigma_\theta$ is in mas.
\label{WL_2star}}
\end{figure}

We should be careful, however.  Most field blends drawn from the LMC
luminosity function are less than a few percent of the brightness of a
$V=19$ star.  A comparably bright star landing in the field of view
could corrupt measurements based upon white-light fringes
\citep{jayadev}.  To illustrate this, we simulate white light SIM
measurements of microlensing events with an equally bright source in
the \fov.  Now, it should be clear that sources within a few fringes
of the target will hopelessly corrupt our measurements.  However,
sources falling outside the coherence envelope have significantly
diminished fringe contrast, and one might hope that their effect would
be minimal.  These hopes are unfounded, as figure \ref{WL_2star}
illustrates.  This figure plots the errors in $\thetaE$ as a function
of source separation (both in mas).  Even sources lying more than 100
mas away can induce 20\% errors in the fit.  A caveat to this
statement is that the assumption of rectangular bandpass overestimates
the effect of distant sources.  The discontinuous jumps in $F(k)$ at
the edges of the bandpass give sidelobes in the white-light fringe
envelope, which would be suppressed with a more smooth decline.
Nevertheless, distant bright sources are clearly cause for concern
when centroiding white light fringes.  Now, space telescope or ground
based AO snapshots could detect such sources.  However, the positions
available from such snapshots are rough by SIM standards, $\sim 20-40$
mas, which is larger than a fringe size and larger than the length
scale of the variations exhibited in Figure \ref{WL_2star}.  Thus,
such high resolution snapshots would not be helpful for fitting
multiple sources.

\begin{figure}
\begin{centering}\plotone{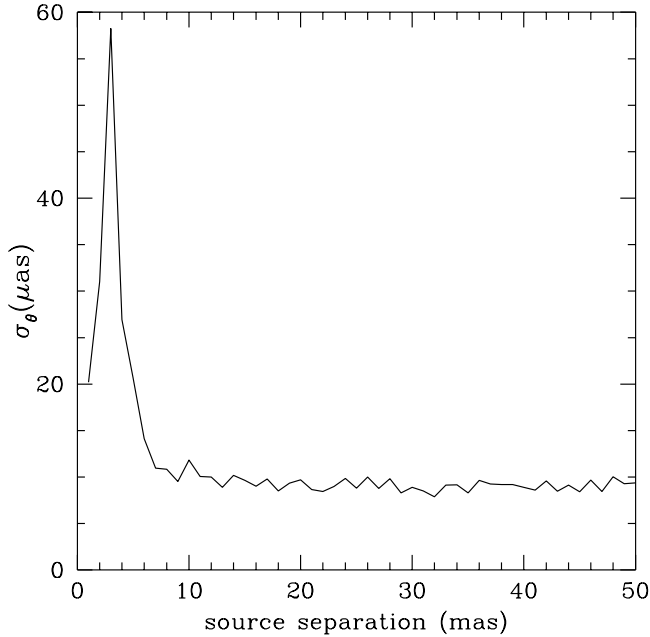}\end{centering}
\caption{Fit residuals for microlensing event with equally bright
blend star and 15 other blends drawn from LMC luminosity function.
The 15 faint LMC blends are placed randomly in \fov, and errors are plotted
as a function of the bright blend position.  Compare this curve with
Figure \ref{2starposition}.
\label{res_2+15}}
\end{figure}

How much better do we do if we have dispersed fringes?
As previous sections showed, bright sources landing outside the central
fringe are easily detected and removed.  Therefore, such
sources will not corrupt SIM microlensing events beyond adding
photon noise, which can be alleviated by integrating longer.  We might
worry about closer sources, however.  Figure \ref{res_2+15} shows
results from the same simulation as above, but for blends within 50
mas, using the channeled spectrum.
The errors are all small ($\sim 3\%$) except for marginally resolved
blends at about 3 mas separation which have 10-20\% errors.  Bright
sources at this distance fall within the central fringe and cannot be
clearly resolved, but are not well enough inside the central fringe 
to be treated as completely unresolved.  In contrast, faint sources at
this distance are effectively unresolved.

This shows the extent to which spectrally dispersed fringes are
useful.  If the spectral decomposition of the fringe pattern is
unavailable or costly to transmit, and we must rely upon white-light
fringes, then high resolution space telescope or AO images will be
required before following up a microlensing event with SIM.  Detection
of a bright field source in such snapshots would not help us remove
the blend's effect, but instead would tell us that such events are
hopelessly corrupted and should not be followed by SIM.  Since bright
events in the LMC are rare, and snapshots would tell us nothing about
possible blends $\lesssim 40$ mas from the target which would also
corrupt SIM measurements, we hope that spectrally dispersed fringes 
will be available in the final SIM design.  High resolution snapshots
may be useful with dispersed fringes as well, since they could be used
to avoid orientations that project bright field stars inside the
target's central fringe.

\vspace{3mm}

\section{Summary}
We have shown that the spectral decomposition of the fringe pattern,
which is often used for group delay estimation, also allows highly
precise astrometry to be performed in crowded fields.  Even when
white-light fringes are hopelessly corrupted, dispersed fringes allow
microarcsecond positions to be measured.  We find that any significant
source outside of the central fringe ($>6$ mas) of the target star may
be detected and removed by fitting the two dimensional (delay {\it
vs.} wavelength) fringe pattern, thereby restoring SIM astrometry to
the precision (few $\mu$as) expected for isolated sources.  We have
further shown that in mildly crowded stellar fields, precise
astrometric imaging is possible using fringes in the channeled
spectrum, with positional precision comparable to that expected for
single star astrometry.  Although the details will depend upon SIM's
precise characteristics, we expect that imaging should be possible for
up to 10 comparably bright sources in the \fov.

We then illustrated our method with the specific example of astrometric
microlensing.  We showed that, although unresolved sources falling
within the central fringe could corrupt individual astrometric
measurements, accurate masses may still be measured by fitting
the entire microlensing event.  In this manner, confusion or blending
errors may be brought below photon noise errors.  For the worst case,
i.e. LMC events which are the most crowded and have relatively faint
sources, 5\% mass measurements are still easily attainable.  We also
considered the effect of exceptionally bright field stars or binary
companions, and showed that
for all separations except about 3 mas, such bright
sources also do not adversely affect SIM astrometry.  Chance
separations of 3 mas, however, can lead to 20\% errors in mass and
distance measurements from microlensing events.

In conclusion, we find that SIM astrometry will not be corrupted by
blending, and extremely precise imaging of mildly crowded fields may
be performed using only two orthogonal baseline orientations, allowing
microarcsecond positional measurements.

\acknowledgments{We thank Andy Boden, Andy Gould, and Andreas Quirrenbach
for many helpful discussions and suggestions, and Jayadev Rajagopal
for helpful discussions, suggestions and for sending us an advance
copy of his paper.  This work was supported in part by the
U.S. Dept. of Energy under grant DEFG0390ER40546.  ND was also
supported in part by the ARCS Foundation.}

\end{document}